\documentclass[aip,pop,reprint,amsmath,amssymb]{revtex4-1}

\usepackage{graphicx}
\renewcommand{\vec}[1]{\boldsymbol{#1}}

\def\aap{{\itshape Astron. \& Astrophys.} }
\def\jgr{{\itshape J. Geophys. Res.} }
\def\apj{{\itshape Astrophys. J.} }
\def\grl{{\itshape Geophys. Res. Lett.} }
\def\pop{{\itshape Phys. Plasmas} }
\def\prl{{\itshape Phys. Rev. Lett.} }

\begin{document}

\title{Loading loss-cone distributions in particle simulations}

\author{Seiji Zenitani}
\affiliation{Space Research Institute, Austrian Academy of Sciences, 8042 Graz, Austria}
\email{seiji.zenitani@oeaw.ac.at}
\affiliation{Research Center for Urban Safety and Security, Kobe University, 1-1 Rokkodai-cho, Nada-ku, Kobe 657-8501, Japan.}
\author{Shin'ya Nakano}
\affiliation{The Institute of Statistical Mathematics, 10-3 Midori-cho, Tachikawa, Tokyo 190-8562, Japan}

\begin{abstract}
Numerical procedures to generate random variates that follow loss-cone velocity distributions in particle simulations are presented. We propose a simple summation algorithm for the Ashour-Abdalla--Kennel-type loss-cone distribution, also known as the subtracted Maxwellian. 
For the Dory-type loss-cone distribution, we use a random variate for the gamma distribution.
Extending earlier algorithms for the kappa and Dory-type distributions,
we construct a novel algorithm to generate a popular form of a kappa loss-cone distribution.
To better express the loss cone, we discuss another family of loss-cone distributions based on the pitch angle. In addition to the acceptance-rejection method, we propose two transformation algorithms that convert an isotropic distribution into a loss-cone distribution. This allows us to generate loss-cone and kappa loss-cone distributions from the Maxwell and kappa distributions.
\end{abstract}

\maketitle

\section{Introduction}

A loss-cone distribution is one of
the most characteristic velocity distribution functions in space, solar, and laboratory plasmas.
The distribution function typically develop
a low-density cavity (a ``loss cone'') in the phase space
in the parallel directions,
when particles are trapped
in the dipole field in planetary magnetospheres,
in a solar magnetic loop, and in plasma mirror devices.
In these sites, electrons and protons with loss-cone distributions
are thought to excite waves via kinetic plasma instabilities.\citep{melrose82,friedel02,aschwanden,treumann06,thorne10,zhang13,xiao15,zhou17,summers21}

To discuss basic properties of plasmas with a loss-cone distribution,
various distribution models have been proposed.
As will be discussed in this paper,
to approximate the loss cone,
\citet{DGH} have modified a bi-Maxwellian
by using the perpendicular velocity $\propto (v_{\perp})^{2j}$
with a loss-cone index $j$.
\citet{AK78} have proposed a subtraction of two bi-Maxwellians.
\citet{summers91} have further incorporated a power-law tail into the Dory-type model.

Over many decades, researchers have been using
particle-in-cell (PIC) simulations or hybrid simulations
to study auroral kilometric radiation in the magnetospheres,\citep{wagner83,prit84}
wave-excitation and subsequent nonlinear wave-particle interaction
in the Earth's inner magnetosphere,\citep{katoh06,hikishima09,shoji11,tao14,golkowski19}
and radio emission in solar active regions.\citep{banacek18,ni20,ning21}
These simulations typically employ
one of the aforementioned loss-cone models
as initial velocity distributions. 
Meanwhile, despite its fundamental role in modeling,
a numerical procedure to load particle velocities
that follow a loss-cone distribution is not well known.
Previous authors may have used some kind of acceptance-rejection methods,
but their procedures were rarely detailed in the literature. 
There is a strong demand for well-documented algorithms,
so that many more modelers join this research field.

The purpose of this article is to provide numerical procedures 
for loading loss-cone velocity distributions by random variates
in particle simulations. 
The rest of this paper is organized as follows.
In Section \ref{sec:AK}, we present
a summation algorithm to generate
Ashour-Abdalla--Kennel type loss-cone distribution,
also known as the subtracted Maxwellian.
In Section \ref{sec:DGH}, we show a simple algorithm
for the Dory-type loss-cone distribution.
In Section \ref{sec:KLC}, we propose
a novel algorithm for the kappa loss-cone (KLC) distribution,
which combine the kappa distribution and the Dory-type distribution.
In Section \ref{sec:PA}, we discuss another family of loss-cone distributions,
based on the pitch angle. We propose transformation algorithms that convert a spherically-symmetric distribution into a loss-cone distribution.
The new method allows us to generate loss-cone and KLC distributions
from the Maxwell and kappa distributions.
In Section \ref{sec:test}, we present simple numerical tests.
Section \ref{sec:discussion} contains discussion and summary.

\section{Subtracted Maxwellian}
\label{sec:AK}

The Ashour-Abdalla--Kennel loss-cone distribution,\citep{AK78}
also known as a subtracted Maxwellian, is defined in Eq.~\eqref{eq:subtracted1}.
Probably this is the most popular form
in theoretical and numerical studies of plasmas with loss-cone distributions.
\begin{align}
\label{eq:subtracted1}
&f(v_{\parallel}, \vec{v}_\perp)
=
\frac{N_0}{\pi^{1/2} \theta_{\parallel}} 
\exp \left( -\frac{v^2_\parallel}{\theta_{\parallel}^2} \right) 
\nonumber \\
&
\times
\frac{1}{\pi \theta_{\perp}^2}
\left\{
\Delta
\exp \left( - \frac{v_{\perp}^2}{\theta_{\perp}^2} \right)
+
\frac{1-\Delta}{1-\beta}
\left[
\exp \left( - \frac{v_{\perp}^2}{\theta_{\perp}^2} \right)
-
\exp \left( - \frac{v_{\perp}^2}{\beta \theta_{\perp}^2} \right)
\right]
\right\}
\end{align}
Here, $N_0$ is the plasma density, and $\theta_{\parallel}$ and $\theta_{\perp}$ are the thermal velocities in the parallel and perpendicular directions.
The two parameters $\Delta \in [0,1], \beta \in [0,1]$ control
the relative density inside the loss cone and the shape of the loss cone.
When $\Delta=0$, the loss cone is nearly empty.
When $\Delta=1$, the loss cone is completely filled and the distribution reproduces an anisotropic bi-Maxwellian.
When $\beta=0$, the distribution is identical to bi-Maxwellian.
As $\beta$ increases, the loss cone becomes apparent.
Since one can easily obtain the background Maxwellian,
we limit our attention to the loss-cone case of $\Delta=0$ in this paper. 
\begin{align}
\label{eq:subtracted2}
f(v_{\parallel}, \vec{v}_\perp)
&=
\frac{N_0}{\pi^{1/2} \theta_{\parallel}}
\exp \left( -\frac{v^2_\parallel}{\theta_{\parallel}^2} \right) 
\nonumber \\
& ~~~
\times
\frac{1}{\pi \theta_{\perp}^2(1-\beta)}
\left\{
\exp \left( - \frac{v_{\perp}^2}{\theta_{\perp}^2} \right)
-
\exp \left( - \frac{v_{\perp}^2}{\beta \theta_{\perp}^2} \right)
\right\}
\\
P_\parallel &= \frac{1}{2} N_0 m\theta_{\parallel}^2, ~~~
P_\perp = \frac{1}{2} N_0 m\theta_{\perp}^2 ({1+\beta})
\end{align}
In the distribution function,
the parallel and perpendicular parts are independent.
Since the parallel part is just a Maxwell distribution,
we focus on the $v_{\perp}$ part.
We consider $v_{\perp 1}\equiv v_{\perp}\cos\varphi$ and $v_{\perp 2}\equiv v_{\perp}\sin\varphi$ in the cylindrical coordinates $(v_{\perp}, \varphi, v_{\parallel})$,
and then we rewrite the $v_{\perp}$ part of Eq.~\eqref{eq:subtracted2}.
\begin{align}
f_{V_{\perp}}(v_{\perp})
=
\frac{2\pi v_{\perp}}{\pi \theta_{\perp}^2(1-\beta)}
\left\{
\exp \left( - \frac{v_{\perp}^2}{\theta_{\perp}^2} \right)
-
\exp \left( - \frac{v_{\perp}^2}{\beta \theta_{\perp}^2} \right)
\right\}
\label{eq:v}
\end{align}
By setting $x \equiv v_{\perp}^2/\theta_{\perp}^2$, we further rewrite
\begin{align}
f_X(x)
&=
\frac{1}{1-\beta}
\left(
\exp \left( - x \right)
-
\exp \left( - \frac{x}{\beta} \right)
\right)
\label{eq:g}
\end{align}

In the $\beta=0$ limit,
we obtain an exponential distribution $f_X(x) = e^{-x}$.
We can easily compute a random variate $x$
that follows the exponential distribution,
by using a uniform random variate $U_1 \sim U(0,1)$.
\begin{align}
x \leftarrow -\log U_1
\label{eq:exp}
\end{align}
In practice, we need to deal with the special case of $U_1=0$,
because it gives a numerical error.
For example, when the uniform variate is drawn from $[0,1)$,
we can use $U_1 \leftarrow (1-U_1)$ instead.
This depends on our choice of libraries and programming languages.

In the $\beta \rightarrow 1$ limit,
Eq.~\eqref{eq:g} is equivalent to a $\beta$-derivative of $e^{-x/\beta}$.
\begin{align}
\lim_{\beta \rightarrow 1} f_X(x)
&=
\lim_{\beta \rightarrow 1}
\left(
\frac{
e^{-x}
-
e^{-x/\beta}
}{1-\beta}
\right)
=
\left.
\left(
\frac{d}{d \beta}e^{-x/\beta}
\right)
\right|_{\beta=1}
=
x
e^{-x}
\label{eq:xex}
\end{align}
This is equivalent to a gamma distribution.
We inform the readers that the gamma distribution
with a shape parameter $k$ and a scale parameter $\lambda$
is defined in $0 \le x$
\begin{align}
{\rm Ga}(x; k, \lambda) =
\frac{ x^{k-1} e^{ -x / \lambda } }{ \Gamma(k) \lambda^{k}}
,~~~~
\int_0^{\infty} {\rm Ga}(x) dx = 1
\end{align}
where $\Gamma(x)$ is the gamma function.
To obtain a gamma distribution with shape $k=2$ and scale $\lambda=1$,
one can use two uniform random variates $U_1$ and $U_2$,
as described by textbooks on random variates.\citep{devroye86,yotsuji10,kroese11}
\begin{align}
x \leftarrow -\log U_1 U_2
\label{eq:gamma}
\end{align}
If $U_1$ and $U_2$ are defined on $0$, one can similarly use $U_1 \leftarrow (1-U_1)$ and $U_2 \leftarrow (1-U_2)$.

\begin{table}
\begin{center}
\caption{Algorithm for the subtracted Maxwellian
\label{table:AK_new}}
\begin{tabular}{l}
\\
\hline
{\bf Algorithm 2}\\
\hline
generate $U_1, U_2, U_3 \sim U(0,1)$ \\
generate $N \sim \mathcal{N}(0,1)$ \\
$x \leftarrow {- \log U_1 - \beta \log U_2}$ \\
$v_{\perp 1}  \leftarrow \theta_{\perp} \sqrt{x} \cos(2\pi U_3)$\\
$v_{\perp 2}  \leftarrow \theta_{\perp} \sqrt{x} \sin(2\pi U_3)$\\
$v_{\parallel}~ \leftarrow \theta_{\parallel} \sqrt{1/2} ~N$\\
{\bf return} $v_{\perp 1}, v_{\perp 2}, v_{\parallel}$\\
\hline
\end{tabular}
\end{center}
\end{table}

For $0 \le \beta \le 1$, we propose to use
two uniform variates $U_1$ and $U_2$ in the following way,
\begin{align}
x \leftarrow -\log U_1 - \beta \log U_2
\label{eq:new}
\end{align}
This covers the two limits of $\beta=0,1$ (Eqs.~\eqref{eq:exp} and \eqref{eq:gamma}).
Here below, we show that this $x$ follows $f_X(x)$ in Eq.~\eqref{eq:g}.
For convenience, we define
\begin{align}
s \equiv-\log U_1,~~t \equiv-\beta\log U_2
\end{align}
It is clear that $s$ follows the exponential distribution,
\begin{align}
s \sim G_{\rm s}(s) = \exp \left( - s \right),~~~s\ge 0
\end{align}
The other variable $t$ also follows the exponential distribution,
but it is rescaled by $\beta$. As a consequence, it follows
\begin{align}
t \sim G_{\rm t}(t) = \frac{1}{\beta} \exp \left( - \frac{t}{\beta} \right),~~~t\ge 0
\end{align}
Then we consider the summation of the two variables,
\begin{align}
x \leftarrow s + t
\end{align}
For specific $x$, we need to consider
all possible combinations of $s$ and $t$.
The distribution function of $x$, $G(x)$, is given by
\begin{align}
G(x)
&= \int_0^x G_{\rm s}(s) \cdot G_{\rm t}(x-s)~ds
\nonumber \\
&
=
\frac{1}{\beta}\int_0^x \exp\left( -\frac{\beta-1}{\beta} s- \frac{x}{\beta} \right) ds
\nonumber \\
&=
\frac{1}{1-\beta}
\left\{
\exp \left( - x \right)
-
\exp \left( - \frac{x}{\beta} \right)
\right\}
= f_X(x)
\end{align}
Thus, the procedure in Eq.~\eqref{eq:new} provides $x$,
which follows the subtracted Maxwellian in Eq.~\eqref{eq:g}.

After obtaining $x$, we can straightforwardly recover
the two components of the perpendicular velocity,
$v_{\perp 1}$ and $v_{\perp 2}$, with help from another uniform variate $U_3$.
\begin{align}
v_{\perp 1} = \theta_{\perp} \sqrt{x} \cos(2\pi U_3),~~~
v_{\perp 2} = \theta_{\perp} \sqrt{x} \sin(2\pi U_3)
\end{align}
The numerical procedure to obtain $v_{\perp 1}$ and $v_{\perp 2}$ 
is presented in Algorithm 2 in Table \ref{table:AK_new}.
For completeness, a procedure to obtain $v_{\parallel}$ is added.
It is just a Maxwellian, and so we can use the \citet{bm58} method or
the built-in procedure for the normal distribution.
Note that it contains a trivial factor of $\sqrt{1/2}$.

\begin{figure}[htbp]
\centering
\includegraphics[width={\columnwidth}]{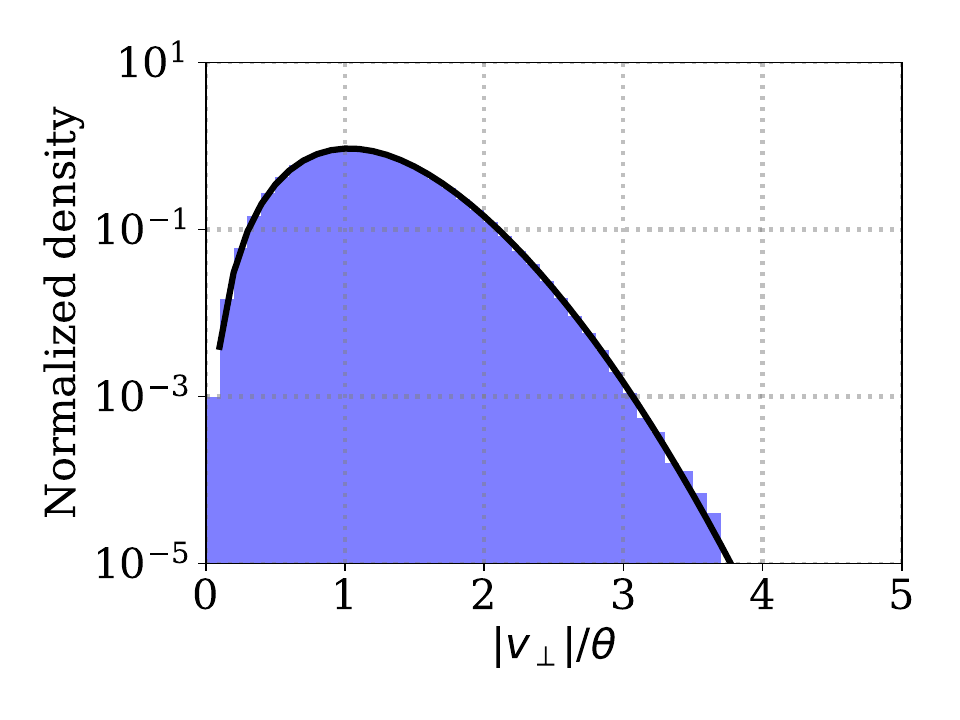}
\caption{
The distribution of $v_{\perp}$ of the subtracted Maxwellian ($\beta=0.5$).
Monte Carlo results with $10^6$ particles (the blue histogram) and the theoretical curve (the black curve) are compared.
\label{fig:AK_hist}}
\end{figure}

\begin{figure}[htbp]
\centering
\includegraphics[width={\columnwidth}]{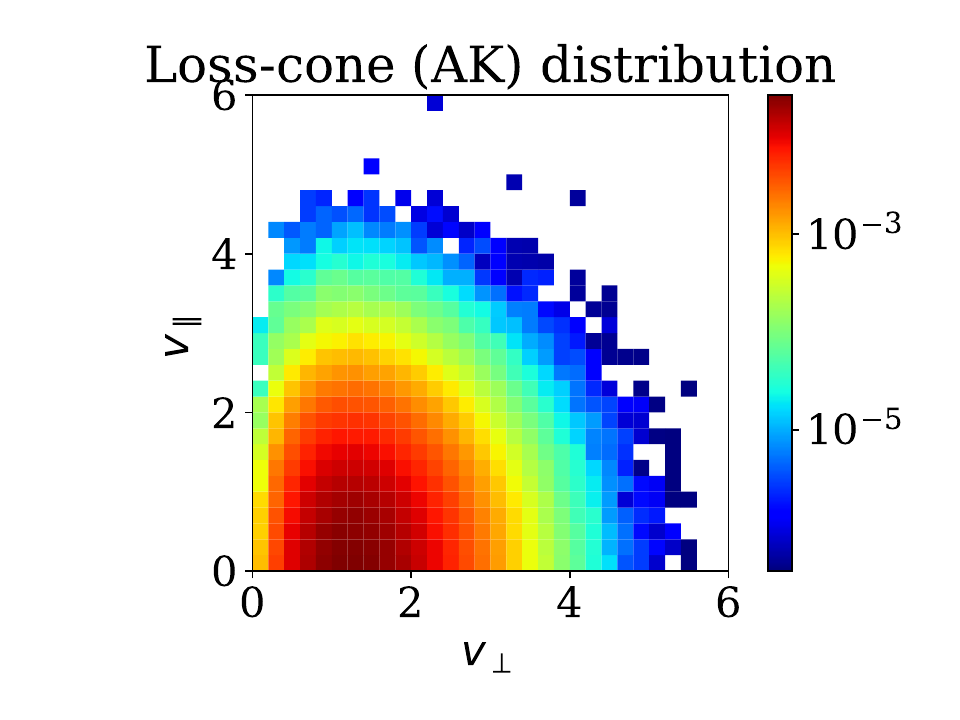}
\caption{
Monte Carlo sampling of the subtracted Maxwellian ($\beta=0.5$ and $\theta=1.5$) with $10^6$ particles.
Phase-space density is presented in the $v_{\perp}$--$v_{\parallel}$ plane.
\label{fig:AK}}
\end{figure}

Using this method, we have numerically generated the loss-cone distribution.
Figure \ref{fig:AK_hist} shows
the distribution of $v_{\perp}$ according to Equation \eqref{eq:v} for $\beta=0.5$.
The blue histogram displays our Monte Carlo results with $10^6$ particles,
and the black curve indicates the analytic solution.
They are in excellent agreement. 
Figure \ref{fig:AK} shows a phase-space density of the distribution
in the $v_{\perp}$--$v_{\parallel}$ plane.
The cell size is $\Delta v = 1/5$.
To enlarge its internal structure, $\theta$ is set to $1.5$.
One can clearly see an empty hole near the $v_{\parallel}$ axis.
There exists a region where $(\partial /\partial v_{\perp})f > 0$.

\section{Dory-type loss-cone distribution}
\label{sec:DGH}

Next we discuss another loss-cone model,
proposed by \citet{DGH}.
This distribution is called
Dory-type loss-cone distribution or
Dory--Guest--Harris (DGH) distribution.
It approximates a loss cone
by using a power of the perpendicular velocity $\propto (v_{\perp}^2)^{j}$.
The loss-cone index $j$ is a non-negative number, $j \ge 0$.
\begin{align}
\label{eq:DGH}
f(v_{\parallel}, \vec{v}_{\perp})
&=
\frac{N_0}{\pi^{3/2}\theta_{\parallel}\theta_{\perp}^2 \Gamma(j+1)}
\left( \frac{v_{\perp}}{\theta_{\perp}} \right)^{2j}
\exp \left( -\frac{v^2_\parallel}{\theta_{\parallel}^2} - \frac{v_{\perp}^2}{\theta_{\perp}^2} \right)
\\
P_\parallel &= \frac{1}{2} N_0 m\theta_{\parallel}^2, ~~~
P_\perp = \frac{1}{2} N_0 m\theta_{\perp}^2 ({1+j})
\end{align}
When the loss-cone index is $j=0$, the distribution is nearly a bi-Maxwellian.
As the index increases, the loss cone becomes apparent.
The subtracted Maxwellian with $\Delta=0$ and $\beta=1$ is identical to
the Dory-type distribution with $j=1$.
We similarly split the distribution into
the parallel and perpendicular parts:
\begin{align}
f(v_{\parallel}, \vec{v}_\perp)
&=
\frac{N_0}{\pi^{1/2}\theta_{\parallel}}
\exp \left( -\frac{v^2_\parallel}{\theta_{\parallel}^2} \right) 
\nonumber \\
&~~~
\times
\frac{1}{\pi\theta_{\perp}^2 \Gamma(j+1)}
\left( \frac{v_{\perp}}{\theta_{\perp}} \right)^{2j}
\exp \left( -\frac{v_{\perp}^2}{\theta_{\perp}^2} \right)
\label{eq:dory2}
\end{align}
Again, we focus on the perpendicular part.
Considering $v_{\perp 1}\equiv v_{\perp}\cos\varphi$ and $v_{\perp 2}\equiv v_{\perp}\sin\varphi$ in the cylindrical coordinates $(v_{\perp}, \varphi, v_{\parallel})$,
we rewrite the perpendicular part as
\begin{align}
f_{V_{\perp}}(v_{\perp})
&=
\frac{2\pi v_{\perp}}{\pi\theta_{\perp}^2 \Gamma(j+1)}
\left( \frac{v_{\perp}^2}{\theta_{\perp}^2} \right)^{j}
\exp \left( -\frac{v_{\perp}^2}{\theta_{\perp}^2} \right)
\end{align}
By setting $x \equiv v_{\perp}^2/\theta_{\perp}^2$,
we find a gamma distribution with a shape parameter $j+1$. 
\begin{align}
f_X(x)
&=
\frac{x^j e^{-x}}{\Gamma(j+1)} = {\rm Ga}(x; j+1, 1)
\label{eq:g3}
\end{align}

There are several algorithms to generate a gamma distribution.
We recommend the readers to consult
textbooks on random variates\citep{devroye86,yotsuji10,kroese11} for detail,
but we quickly outline the gamma-distribution generators here.
Let us consider the gamma distribution with shape $k$ and scale $\lambda$, i.e. $\sim {\rm Ga}(k,\lambda)$.
When $k$ is a integer, a gamma variate $x$ can be drawn
by using multiple uniform random variates
$Y_i \sim U(0,1)$ ($i=1, 2, \dots, k$) in the following way,
\begin{align}
x \leftarrow
\lambda \sum_{i=1}^{k} \left( -\log Y_i \right) = - \lambda \log \left( \prod_{i=1}^{k} Y_i \right)
\label{eq:gamma_int}
\end{align}
When $k$ is a half integer ($1/2, 3/2, \cdots$),
in addition to uniform variates $Y_i \sim U(0,1)$ ($i=1, 2, \dots k-1/2$),
we use one normal variate $n_1 \sim \mathcal{N}(0,1)$,
\begin{align}
x \leftarrow
- \lambda \log \left( \prod_{i=1}^{k-1/2} Y_{i} \right) + \frac{\lambda}{2} n_1^2
.
\label{eq:gamma_half}
\end{align}
When $k>1$ is non-integer, \citet{mt00}'s method is useful,
as detailed by many textbooks.
One can also use it for integer and non-integer cases, because it is known to be faster.

In the case of Eq.~\eqref{eq:g3},
we can generate a gamma variate $x \sim {\rm Ga}(j+1,1)$
by using Eq.~\eqref{eq:gamma_int}.
Then we similarly obtain $v_{\perp 1}$ and $v_{\perp 2}$. 
The procedure is outlined in Algorithm 3 in Table \ref{table:DGH}.
The gamma generator for integer $j$ is written in the form of comments,
in case we employ other algorithms for non-integer $j$.
When $j=0$, the perpendicular part recovers the \citet{bm58} method to generate a Maxwellian.

\begin{figure}[htbp]
\centering
\includegraphics[width={\columnwidth}]{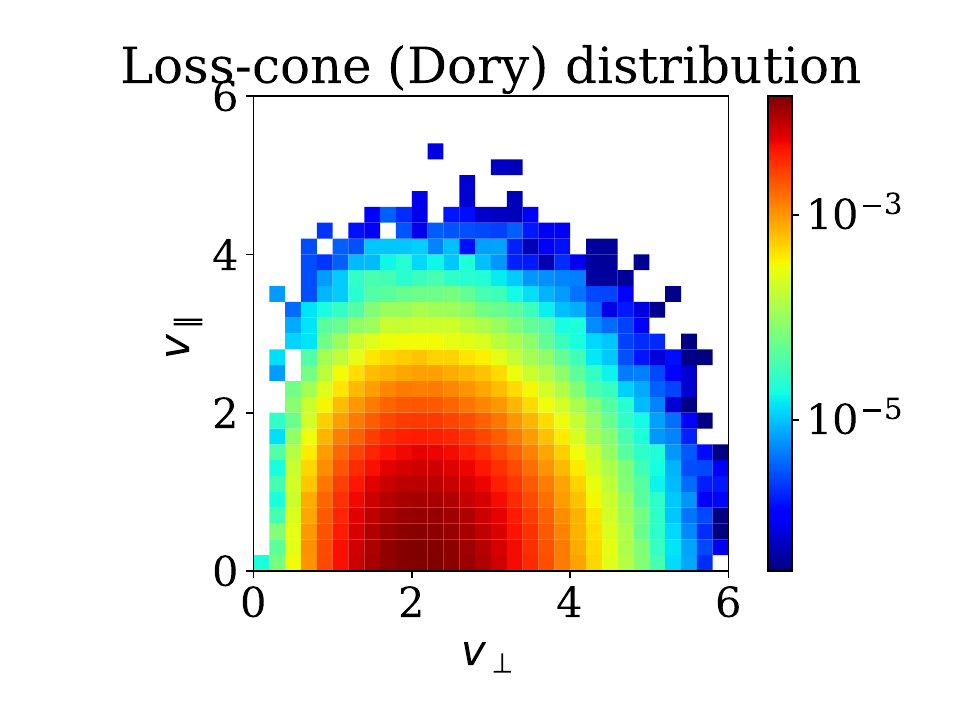}
\caption{
Monte Carlo sampling of the Dory-type loss-cone distribution ($\theta=1.5$ and $j=2$) with $10^6$ particles.
Phase-space density is presented in the $v_{\perp}$--$v_{\parallel}$ plane.
\label{fig:DGH}}
\end{figure}

We have numerically generated the Dory-type distribution of $10^6$ particles,
according to Equation \eqref{eq:DGH} with $j=2$.
Fig.~\ref{fig:DGH} shows the phase-space density
in the $v_{\perp}$--$v_{\parallel}$ plane,
in the same format as Fig.~\ref{fig:AK}.
To emphasize the internal structure, $\theta$ is set to $1.5$.
There is a vertical hole with a steep density gradient along the $v_{\parallel}$ axis.

\begin{table}
\begin{center}
\caption{Algorithm for the Dory-type loss-cone distribution
\label{table:DGH}}
\begin{tabular}{l}
\\
\hline
{\bf Algorithm 3}\\
\hline
generate $X \sim {\rm Ga}(j+1,1)$\\
// generate $Y_1, \cdots, Y_{j+1} \sim U(0,1)$ \\
// $X \leftarrow -\log \big(\prod_{k=1}^{j+1} Y_{k} \big)$\\
generate $U \sim U(0, 1)$ \\
generate $N \sim \mathcal{N}(0,1)$ \\
$v_{\perp 1} \leftarrow \theta_{\perp} \sqrt{ X } \cos(2\pi U)$ \\
$v_{\perp 2} \leftarrow \theta_{\perp} \sqrt{ X } \sin(2\pi U)$ \\
$v_{\parallel}~ \leftarrow \theta_{\parallel} \sqrt{1/2} ~N$\\
{\bf return} $v_{\perp 1}, v_{\perp 2}, v_{\parallel}$\\
\hline
\end{tabular}
\end{center}
\end{table}

\section{Kappa loss-cone distribution}
\label{sec:KLC}

The kappa distribution\citep{vas68,olbert68} extends the Maxwell distribution with a power-law tail in the high-energy part, and it has drawn huge attention in space physics.\citep{kappa}
Recognizing the importance of the kappa distribution,
\citet{summers91} have proposed a hybrid distribution of the kappa distribution and
the Dory-type loss-cone distribution.
It was sometimes called the (generalized) Lorentzian loss-cone distribution, while
it is popularly referred to as the kappa loss-cone (KLC) distribution.
In this paper, we occasionally call it the Summers-type KLC distribution to
distinguish it from another variant in Section \ref{sec:application}.
Its mathematical form is given by \citet{summers91,summers95}:
\begin{align}
\label{eq:KLC}
f(\vec{v})
&= \frac{N_0}{\pi^{3/2}\theta_{\parallel}\theta_{\perp}^2\kappa^{j+3/2}}
\frac{\Gamma(\kappa+j+1)}{\Gamma(j+1)\Gamma(\kappa-1/2)}
\nonumber \\
&~~~
\times
\left( \frac{v_{\perp}}{\theta_{\perp}} \right)^{2j}
\Big( 1 + \frac{ {v_{\parallel}}^2 }{\kappa \theta_{\parallel}^2}  + \frac{ {v_{\perp}}^2 }{\kappa \theta_{\perp}^2} \Big)^{-(\kappa+j+1)} \\
P_{\parallel} &= \frac{\kappa}{2\kappa-3} N_0 m\theta_{\parallel}^2, ~~~
P_{\perp} = \frac{\kappa}{2\kappa-3} N_0 m\theta_{\perp}^2 (1+j)
\end{align}
Here, $\kappa$ ($> 3/2$) is the kappa index and $j$ ($\ge 0$) is the loss-cone index.
When $j=0$, the distribution recovers a standard (bi-)kappa distribution.
When $j>0$, it develops a loss-cone.

The KLC distribution has nice mathematical properties.
For example, in the isotropic case of $\theta_{\parallel} = \theta_{\perp} = \theta$,
Eq.~\eqref{eq:KLC} yields
\begin{align}
f(\vec{v})
&\propto \Big( 1 + \frac{ {v}^2 }{\kappa \theta^2} \Big)^{-(\kappa+1)}
\left(
\dfrac{ 
\frac{v_{\perp}^2}{\kappa\theta^2}
}
{
1 + \frac{ {v}^2 }{\kappa \theta^2}
}
\right)^j
\nonumber \\
&~~~
\approx
\Big( {\rm ~\cdots~} \Big)^{-(\kappa+1)}
\Big(
\sin\alpha
\Big)^{2j}
\end{align}
Along with a power-law tail with an index of $\kappa+1$,
one can see that
the loss-cone is approximated by the pitch angle $\alpha$
in the $v^2 \gg \kappa\theta^2$ range,
as was done in an earlier work.\citep{kennel66}

\begin{table}
\begin{center}
\caption{Algorithm for the Summers-type kappa loss-cone (KLC) distribution
\label{table:KLC}}
\begin{tabular}{l}
\\
\hline
{\bf Algorithm 4}\\
\hline
generate $N \sim \mathcal{N}(0,1)$ \\
generate $Y \sim {\rm Ga}(\kappa-1/2,2)$ \\
generate $X \sim {\rm Ga}(j+1,2)$ \\
generate $U \sim U(0, 1)$ \\
$v_{\perp 1} \leftarrow \theta_{\perp} \sqrt{ \kappa X } \cos(2\pi U) / \sqrt{Y}$ \\
$v_{\perp 2} \leftarrow \theta_{\perp} \sqrt{ \kappa X } \sin(2\pi U) / \sqrt{Y}$ \\
$v_{\parallel} ~ \leftarrow \theta_{\parallel} \sqrt{ \kappa } N / \sqrt{Y}$ \\
{\bf return} $v_{\perp 1}, v_{\perp 2}, v_{\parallel}$\\
\hline
\end{tabular}
\end{center}
\end{table}

A numerical procedure for the Dory-type distribution was discussed in Section \ref{sec:DGH}.
Procedures for the kappa distribution are also known.\citep{abdul15,zeni22}
For detail, the readers may wish to consult Section III in \citet{zeni22} (Hereafter referred to as ZN22).
Combining the procedures for the Dory distribution (Table \ref{table:DGH}) and for the kappa distribution (Table I in ZN22),
we propose a novel procedure for the KLC distribution in Table \ref{table:KLC}.
This algorithm uses two gamma-distributed variates.
Their shape parameters can be an integer, a half-integer, and non-integer.
In any case, we can obtain the random variates
by using gamma generators in Section \ref{sec:DGH}.
Below, we gives a formal proof that
the new procedure generates the KLC distribution according to Eq.~\eqref{eq:KLC}.

We consider four independent random variates $X$, $N$, $Y$, and $U$.
The first variate $X$ follows a gamma distribution
with the shape parameter $j+1$ and the scale parameter 2,
$X \sim {\rm Ga}(j+1, 2)$.
The second follows a normal distribution,
$N \sim \mathcal{N}(0,1)$.
The third follows a chi-squared distribution
with $2\kappa-1$ degrees of freedom.
This chi-squared distribution is equivalent to
the gamma distribution of $Y \sim {\rm Ga}(\kappa-1/2,2)$.
The last variate follows the uniform distribution, $U \sim U(0,1)$.
Using the four variates, we define the following four variables.
\begin{align}
V_{\perp 1} &= \theta_{\perp} \sqrt{ \kappa X } \cos(2\pi U) / \sqrt{Y} \\
V_{\perp 2} &= \theta_{\perp} \sqrt{ \kappa X } \sin(2\pi U) / \sqrt{Y} \\
V_{\parallel} &= \theta_{\parallel} \sqrt{ \kappa } N / \sqrt{Y} \\
Z &= X + N^2 + Y
\label{eq:Z}
\end{align}
We immediately obtain
\begin{align}
\label{eq:X}
X & 
= \frac{V_{\perp}^2}{\kappa\theta_{\perp}^2} Y 
= \frac{V_{\perp}^2}{\kappa\theta_{\perp}^2} Z \left( 1 + \frac{V_{\parallel}^2}{\kappa\theta_{\parallel}^2} + \frac{V_{\perp}^2}{\kappa\theta_{\perp}^2} \right)^{-1} \\
\label{eq:N}
N &= \frac{V_{\parallel}}{\sqrt{\kappa}\theta_{\parallel}} \sqrt{Y}
= \frac{V_{\parallel}}{\sqrt{\kappa}\theta_{\parallel}} \sqrt{Z} 
\left( 1 + \frac{V_{\parallel}^2}{\kappa\theta_{\parallel}^2} + \frac{V_{\perp}^2}{\kappa\theta_{\perp}^2} \right)^{-1/2} \\
\label{eq:Y}
Y &= Z \left( 1 + \frac{V_{\parallel}^2}{\kappa\theta_{\parallel}^2} + \frac{V_{\perp}^2}{\kappa\theta_{\perp}^2} \right)^{-1}
\end{align}
With help from a chain rule, we calculate the following Jacobian
\begin{align}
\label{eq:Jacobian}
\left\| \frac{\partial(X, N, Y)}{\partial(V_{\perp}, V_{\parallel}, Z)} \right\|
&=
\left\| \frac{\partial(X, N, Y)}{\partial(V_{\perp}, V_{\parallel}, Y)} \right\|
\cdot
\left\| \frac{\partial(V_{\perp}, V_{\parallel}, Y)}{\partial(V_{\perp}, V_{\parallel}, Z)} \right\|
\nonumber \\
&
=
\frac{ 2 V_{\perp} Y^{3/2} }{\kappa^{3/2}\theta_{\parallel}\theta_{\perp}^2}
\cdot \left| \frac{\partial Y}{\partial Z} \right|
\nonumber \\
&
=
\frac{ 2 V_{\perp} Z^{3/2} }{\kappa^{3/2}\theta_{\parallel}\theta_{\perp}^2}
\left( 1 + \frac{V_{\parallel}^2}{\kappa\theta_{\parallel}^2} + \frac{V_{\perp}^2}{\kappa\theta_{\perp}^2} \right)^{-5/2}
\end{align}

We consider the joint probability distribution function of $X$, $N$, and $Y$.
Since the three variates are independent,
the function is a product of the gamma distribution of $x$, the normal distribution of $n$, and the gamma distribution of $y$,
\begin{widetext}
\begin{align}
f_{X,N,Y}(x,n,y) = 
\left( \frac{ x^{j} e^{ -x / 2 } }{ \Gamma(j+1) 2^{j+1}} \right)
\times
\left(\frac{1}{\sqrt{2\pi}}e^{-n^2/2} \right)
\times
\left( \frac{ y^{\kappa-3/2} e^{ -y / 2 } }{ \Gamma(\kappa-1/2) 2^{\kappa-1/2}} \right)
\label{eq:fXNY}
\end{align}
Using Eqs.~\eqref{eq:X}--\eqref{eq:Jacobian}, we translate Eq.~\eqref{eq:fXNY} into
\begin{align}
f_{V_{\perp},V_{\parallel},Z}(v_{\perp},v_{\parallel},z)
&= 
\frac{ z^{j+\kappa-3/2} e^{ -z / 2 } }
{ \sqrt{\pi}\Gamma(j+1) \Gamma(\kappa-1/2) 2^{j+\kappa+1}}
\left( 
\frac{v_{\perp}^2}{\kappa\theta_{\perp}^2}
\right)^j
\left( 1 + \frac{v_{\parallel}^2}{\kappa\theta_{\parallel}^2} + \frac{v_{\perp}^2}{\kappa\theta_{\perp}^2} \right)^{-j-\kappa+3/2}
\left\| \frac{\partial(X, N, Y)}{\partial(V_{\perp}, V_{\parallel}, Z)} \right\|
\nonumber \\
&=
\frac{ 2 \pi v_{\perp}  }{\kappa^{3/2}\theta_{\parallel}\theta_{\perp}^2}
\frac{ z^{j+\kappa} e^{ -z / 2 } }
{ \pi^{3/2}\Gamma(j+1) \Gamma(\kappa-1/2) 2^{j+\kappa+1}}
\left( 
\frac{v_{\perp}^2}{\kappa\theta_{\perp}^2}
\right)^j
\left( 1 + \frac{v_{\parallel}^2}{\kappa\theta_{\parallel}^2} + \frac{v_{\perp}^2}{\kappa\theta_{\perp}^2} \right)^{-(j+\kappa+1)}
\nonumber \\
&=
\left( \frac{ z^{j+\kappa} e^{-z/2}}{ 2^{j+\kappa+1}\Gamma(j+\kappa+1)} \right)
\nonumber \\
& ~~~\times
\left\{
\frac{ 1 }{ {\pi}^{3/2}\theta_{\parallel}\theta_{\perp}^2 \kappa^{j+3/2}}
\frac{ \Gamma(j+\kappa+1) }{ \Gamma(j+1)\Gamma(\kappa-1/2) }
\left( \frac{v_{\perp}}{\theta_{\perp}} \right)^{2j}
\left( 1 + \frac{v_{\parallel}^2}{\kappa\theta_{\parallel}^2} + \frac{v_{\perp}^2}{\kappa\theta_{\perp}^2} \right)^{-(j+\kappa+1)}
2\pi v_{\perp}
\right\}
\label{eq:proof}
\end{align}
\end{widetext}
This tells us that
the variable $Z$ follows the gamma distribution with shape $j+\kappa+1$ and scale $2$,
$Z \sim {\rm Ga}(j+\kappa+1,2)$,
and the other variables $V_{\perp}$ and $V_{\parallel}$ are distributed
by the KLC distribution (Eq.~\eqref{eq:KLC}),
with trivial translation of $V_{\perp} \rightarrow (V_{\perp 1}, V_{\perp 2})$.
Eq.~\eqref{eq:proof} also indicates that the two distributions are independent.

Mathematically, it is reasonable that $Z$ follows the gamma distribution.
The square of the normal distribution $N$ provides a chi-squared distribution with one degree of freedom,
which is equivalent to the gamma distribution ${\rm Ga}(1/2,2)$.
This means that $Z$ is a sum of the three gamma distributions with the same scale parameter (Eq.~\eqref{eq:Z}).
In such a case, it is known that their sum follows a gamma distribution with the summed shape parameter:
$Z \sim {\rm Ga}\left((\kappa-\frac{1}{2})+\frac{1}{2}+(j+1),2\right)$,
in agreement with Eq.~\eqref{eq:proof}.

\begin{figure}[htbp]
\centering
\includegraphics[width={\columnwidth}]{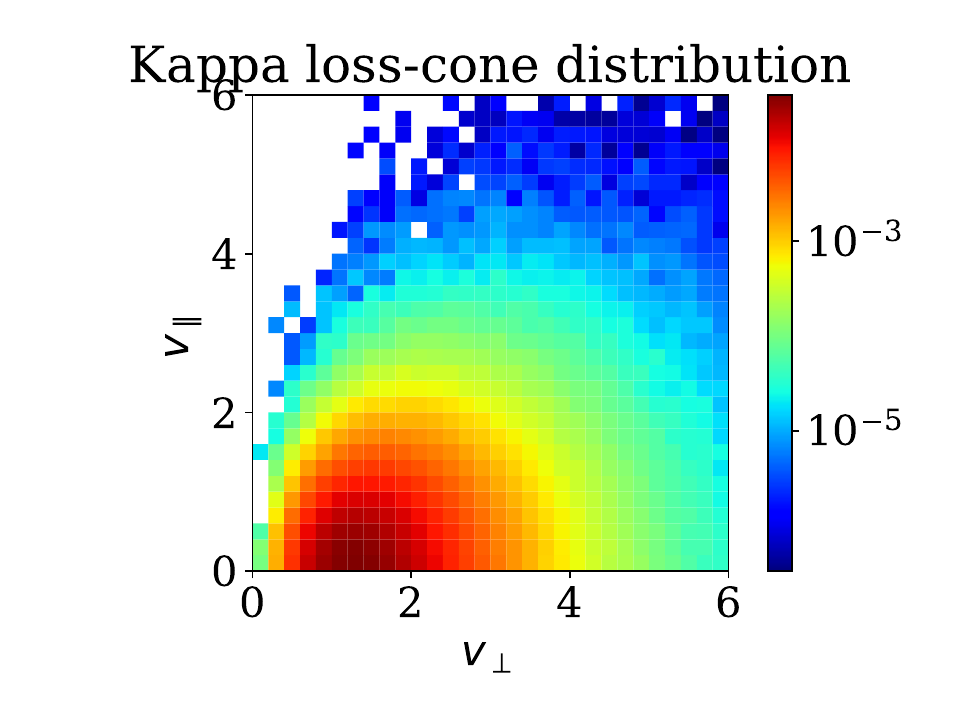}
\caption{
Monte Carlo sampling of the KLC distribution ($\theta=1.0$, $j=2$, and $\kappa=3.5$) with $10^6$ particles.
Phase-space density is shown in $v_{\perp}$--$v_{\parallel}$.
\label{fig:KLC}}
\end{figure}

Using Algorithm 4 in Table \ref{table:KLC},
we have numerically generated the KLC distribution with $10^6$ particles.
The parameters are set to $\theta_{\perp}=\theta_{\parallel}=1.0$, $j=2$ and $\kappa=3.5$.
Figure \ref{fig:KLC} shows the phase-space density in the $v_{\perp}$--$v_{\parallel}$ space
in the same format.
There is a vertical hole near the $v_{\parallel}$ axis, and
plasmas are widely spread over the velocity space. 
In the horizontal direction (at $v_{\parallel}=0$), the phase-space density
decays like $\propto v_{\perp}^{-2(\kappa+1)} = v_{\perp}^{-9.0}$.
As $v_{\perp}$ increases, it drops slower than
in the other loss-cone distributions ($\propto \exp[-(v_{\perp}^2/\theta_\perp^2)]$).
The numerical results are in excellent agreement with the analytic solution,
as will be shown later.


\section{Pitch-angle-type loss-cone distributions}
\label{sec:PA}

\subsection{Acceptance-rejection method}
\label{sec:accept}

In the Dory and KLC distributions, the loss cone is modeled by
a power of the perpendicular velocity, $\propto (v_{\perp})^{2j}$.
Consequently, the ``loss cone'' often looks like a vertical hole,
as evident in Figures \ref{fig:DGH} and \ref{fig:KLC}.
In this section, we consider another loss-cone model,
whose phase-space density is modeled by the pitch angle $\alpha$,
i.e.,
\begin{align}
\propto (\sin\alpha)^{2j} = \left( \frac{v_{\perp}}{v} \right)^{2j}
\label{eq:sin2alpha}
\end{align}
as considered in an earlier work.\citep{kennel66}
We call it the pitch-angle-type (PA-type) loss-cone distribution.

The easiest way to generate a PA-type distribution is
to generate an isotropic distribution, which we call the base distribution,
and then to employ the acceptance-rejection method.
Using a uniform variate $U \sim U(0,1)$,
we accept the particle when the following condition is met,
\begin{align}
U < \left( \frac{v_{\perp}}{v} \right)^{2j}
\end{align}
If this condition is not met, we reject the particle,
and then we regenerate the random variate. 
Then we can straightforwardly obtain the loss-cone distribution,
based on Eq.~\eqref{eq:sin2alpha}.

\begin{figure}[htbp]
\centering
\includegraphics[width={0.5\columnwidth}]{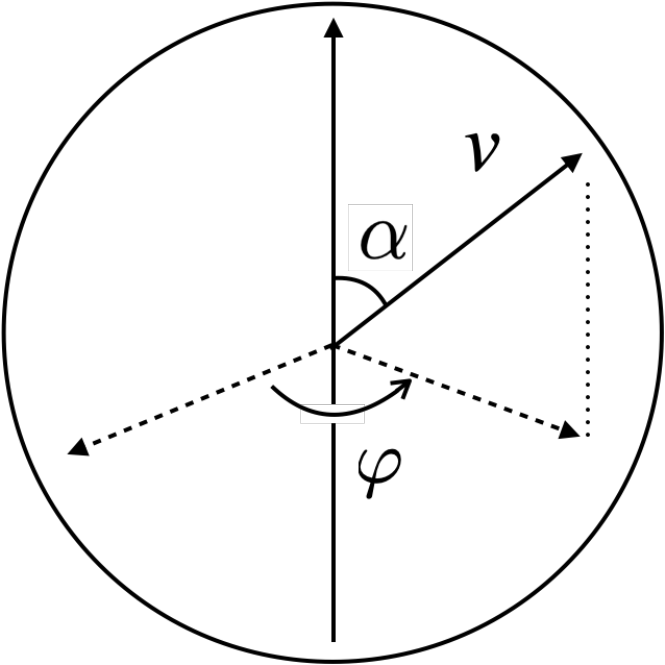}
\caption{
Our spherical coordinates: velocity $v$ ($|\vec{v}|)$, pitch angle $\alpha$, and azimuthal angle $\varphi$.
\label{fig:sphere}}
\end{figure}

\begin{figure}[htbp]
\centering
\includegraphics[width={\columnwidth}]{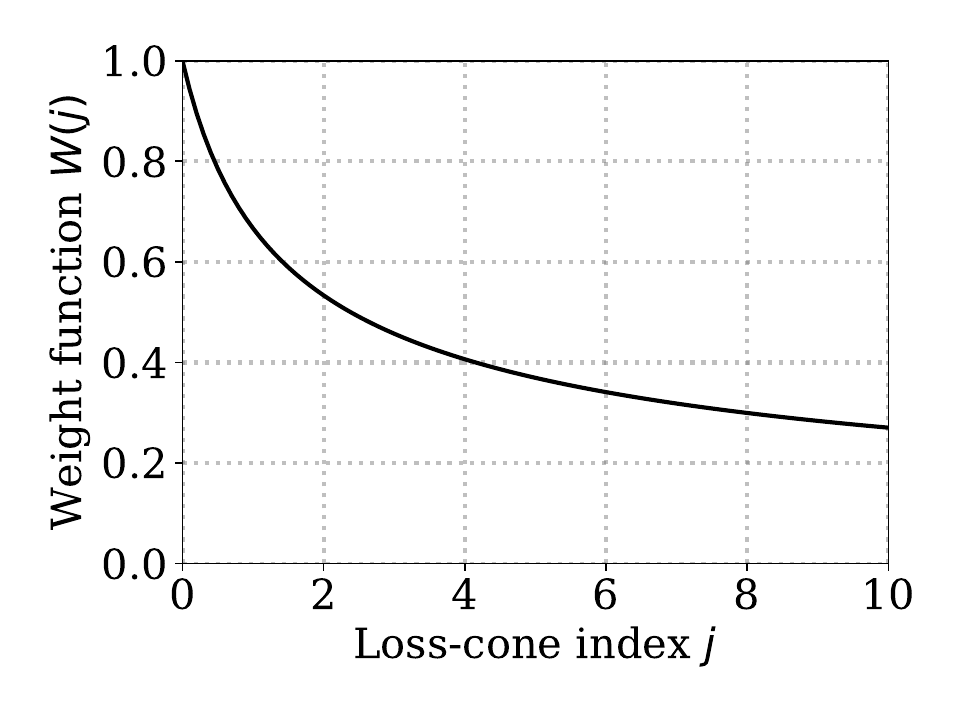}
\caption{
The weight function $W$ as a function of the loss-cone index $j$
(Eq.~\eqref{eq:W})
\label{fig:weight}}
\end{figure}

\begin{widetext}
Let us evaluate the acceptance efficiency of this rejection method.
We consider the spherical coordinates $(v, \alpha, \varphi)$
as illustrated in Fig.~\ref{fig:sphere}.
Since the base distribution $f_0$ is spherically symmetric, it satisfies
\begin{align}
f_0 (\vec{v})~d^3v
=
f_0 (v, \alpha, \varphi) ~v^2 \sin \alpha ~dv d\alpha d\varphi
=
f_0 (v) ~v^2 \sin \alpha ~dv d\alpha d\varphi
\end{align}
We further impose an additional weight of Eq.~\eqref{eq:sin2alpha},
i.e., $f(v) \propto f_0(v)~(\sin \alpha)^{2j}$.
The total weight $W(j)$ is estimated by
\begin{align}
W(j)
&\equiv
\dfrac{\iiint
f_0(\vec{v})~(\sin \alpha)^{2j}
d^3v
}{
\iiint
f_0(\vec{v})~
d^3v
}
=
\dfrac{
\int_0^\infty dv
\int_0^{\pi} d\alpha
\int_0^{2\pi} d\varphi
~
\Big\{
f_0(v)~v^2 (\sin \alpha)^{2j+1}
\Big\}
}{
\int_0^\infty dv
\int_0^{\pi} d\alpha
\int_0^{2\pi} d\varphi
~
\Big\{
f_0(v)~v^2 (\sin \alpha)
\Big\}
}
\nonumber
\\
&=
\dfrac{
4\pi
\Big(
\int_0^\infty
v^2 f_0(v)
~dv
\Big)
\Big(
\int_0^{\pi/2}
\left( \sin \alpha \right)^{2j+1}
d\alpha
\Big)
}{
4\pi
\int_0^\infty
v^2 f_0(v)
~dv
}
=
\int_0^{\pi/2} \left( \sin \alpha \right)^{2j+1} d\alpha
\end{align}
By setting $x \equiv \cos^{2}\alpha$, we obtain 
\begin{align}
W(j)
&=
\frac{1}{2} \int_0^1  (1-x)^{j} x^{-1/2} dx
=
\dfrac{B(1/2,j+1)}{2}
\int_0^1 {\rm Beta}\left(x; \frac{1}{2},j+1 \right) dx
\nonumber \\
&=
\dfrac{\sqrt{\pi}}{2}
\cdot
\dfrac{\Gamma(j+1)}{\Gamma(j+3/2)}
\label{eq:W}
\end{align}
\end{widetext}
where $B(\alpha,\beta)$ is the beta function and
${\rm Beta}(x;\alpha,\beta)$ is the probability density function of the beta distribution.
Since Eq.~\eqref{eq:W} is normalized, it immediately gives
the total acceptance efficiency of the acceptance-rejection method.
As shown in Fig.~\ref{fig:weight},
Eq.~\eqref{eq:W} monotonically decreases from $1$ at $j=0$ to $0$ at $j\rightarrow \infty$.
The efficiency decays slowly, and so
the acceptance-rejection method would be useful for small $j$.

\subsection{Loss-cone transform methods}
\label{sec:LCT}

Recognizing that $x=\cos^2\alpha$ is distributed under the beta distribution in Eq.~\eqref{eq:W},
we propose an algorithm to transform an isotropic distribution into a PA-type loss-cone distribution. 
We first obtain a radial profile of the base distribution, $f_0(v) 4\pi v^2$.
Then we scatter the direction of particle velocity, by using a beta-distributed variate.
This method does not reject any particles. 

A random variate following the beta distribution can be generated
from two gamma-distributed variates,\citep{devroye86,yotsuji10,kroese11}
\begin{align}
\dfrac{X_{\rm Ga(1/2,\lambda)}}{X_{\rm Ga(1/2,\lambda)} + X_{\rm Ga(j+1,\lambda)}} \sim {\rm Beta}(x; 1/2,j+1)
\label{eq:beta}
\end{align}
where $X_{\rm Ga(k,\lambda)}$ is a random variate
following the gamma distribution with shape $k$ and scale $\lambda$.
This $\lambda$ is an arbitrary number, and we set $\lambda=2$ for convenience.
Mathematically, ${\rm Ga(1/2,2)}$ is equivalent to the chi-squared distribution with 1 degree of freedom, which can be obtained from a square of the normal distribution. 
Using a normal variate $N \sim \mathcal{N}(0,1)$ and considering $0 \le \alpha \le \pi$,
we can generate the distribution of the cosine of the pitch angle
\begin{align}
\cos\alpha \sim \dfrac{N}{\sqrt{ N^2 + X_{\rm Ga(j+1,2)}}},
\label{eq:cos}
\end{align}
For given $|v|$, we obtain $v_{\parallel} = |v|\cos\alpha$ and $v_{\perp} = |v| \sin\alpha = |v| (1-\cos^2\alpha)^{1/2}$, and then
we can further obtain $v_{\perp 1}$ and $v_{\perp 2}$ by randomly setting the azimuthal angle. 
The numerical procedure is summarized in Algorithm 5.1 in Table \ref{table:transform}.
We call it the loss-cone transform method.
Applications to the Maxwell and kappa distribution will be presented in Section \ref{sec:application}.

\begin{table}
\centering
\caption{Loss-cone transform algorithms
\label{table:transform}}
\begin{minipage}{0.45\textwidth}
\begin{tabular}{l}
\hline
{\bf Algorithm 5.1: loss-cone transform}\\
$v \mapsto {v}_{\perp 1}, {v}_{\perp 2}, {v}_\parallel$\\
\hline
generate $N \sim \mathcal{N}(0,1)$ \\
generate $X \sim {\rm Ga}(j+1,2)$ \\
generate $U \sim U(0,1)$ \\
${v}_{\perp 1} \leftarrow v ~\sqrt{\dfrac{X}{N^2+X}} \cos(2\pi U)$ \\
${v}_{\perp 2} \leftarrow v ~\sqrt{\dfrac{X}{N^2+X}} \sin(2\pi U)$ \\
${v}_\parallel~ \leftarrow v ~\dfrac{N}{\sqrt{N^2+X} }$ \\
{\bf return} ${v}_{\perp 1}, {v}_{\perp 2}, {v}_\parallel$\\
\hline
\\
\end{tabular}
\end{minipage}
\begin{minipage}{0.45\textwidth}
\begin{tabular}{l}
\hline
{\bf Algorithm 5.2: latitude transform}\\
${v}_{\perp 1}, {v}_{\perp 2}, {v}_\parallel \mapsto \bar{v}_{\perp 1}, \bar{v}_{\perp 2}, \bar{v}_\parallel$\\
\hline
$v^2_\perp \leftarrow v_{\perp 1}^2 + v_{\perp 2}^2$ \\
$v \leftarrow \sqrt{v_{\perp}^2 + v_{\parallel}^2} $ \\
$\bar{v}_\parallel \leftarrow v~C^{-1} \left( {v_{\parallel}}/{v};j \right)$ \\
$\bar{v}_{\perp 1} \leftarrow \sqrt{v^2-\bar{v}_\parallel^2}~\dfrac{v_{\perp 1}}{v_{\perp}}$ \\
$\bar{v}_{\perp 2} \leftarrow \sqrt{v^2-\bar{v}_\parallel^2}~\dfrac{v_{\perp 2}}{v_{\perp}}$ \\
{\bf return} $\bar{v}_{\perp 1}, \bar{v}_{\perp 2}, \bar{v}_\parallel$\\
\hline
\\
\\
\end{tabular}
\end{minipage}
\end{table}

In the loss-cone transform algorithm,
particle velocities are randomly scattered into all the directions.
Here we propose another transform algorithm that only adjusts the pitch angle
but preserves the azimuthal angle. 
We assume $j$ to be an integer in this subsection. 
We consider a cumulative distribution function (CDF) of $x$.
The CDF of the beta distribution is given by
the regularized incomplete beta function, $I_x$.
\begin{align}
C(x;j)
&= I_x(1/2,j+1) = \frac{ {\rm Beta}(x;1/2,j+1) }{B(1/2,j+1)}
\nonumber \\
&= \frac{ 2\sqrt{x}~{}_2F_1(1/2,-j;3/2; x) }{B(1/2,j+1)}
\end{align}
Here, we have used a relation in \citet{AS72} and
${}_2F_1$ is the hypergeometric function.
We further set the cosine to $u = \cos \alpha$.
We temporarily limit our attention to $0 \le u \le 1$, and then
obtain a CDF
\begin{align}
C(u;j) = \frac{ 2u~{}_2F_1(1/2,-j;3/2; u^2) }{B(1/2,j+1)}
\end{align}
This gives
\begin{align}
C(u;0) &= u\\
C(u;1) &= \frac{3}{2} u -\frac{1}{2}u^3 \\
C(u;2) &= \frac{15}{8} u - \frac{5}{4}u^3 + \frac{3}{8}u^5 \\
C(u;3) &= \cdots
\end{align}
These are non-injective functions, but they are monotonic in $-1 \le u \le 1$.
Thus we can define their inverse functions, $C^{-1}(x;j)$ in this range.
When we transform an isotropic distribution into a loss-cone distribution,
we map the cosine latitude $u_0 \mapsto u_j$ such that
\begin{align}
C(u_0;0) &= u_0 = C(u_j;j)
.
\end{align}
Thus, we obtain $u_j = C^{-1}(u_0;j)$ via the inverse function,
which also works for the negative case of $-1 \le u < 0$.
After modifying the latitude, then we adjust $\sin \alpha$ accordingly.
These procedures are summarized in Algorithm 5.2 in Table \ref{table:transform}.
We call this method the latitude transform method.
The azimuthal angle in the velocity space is conserved.
The inverse function can be calculated
by interpolating and looking up a numerical table of the CDF.

\begin{table}
\caption{Algorithms for the PA-type distributions.
\label{table:LC-PA}}
\begin{minipage}{0.45\textwidth}
\begin{tabular}{l}
\hline
{\bf Algorithm 5.3: Loss-cone distribution}\\
\hline
generate $N \sim \mathcal{N}(0,1)$ \\
generate $X_1 \sim {\rm Ga}(3/2,1)$ \\
generate $X_2 \sim {\rm Ga}(j+1,2)$ \\
generate $U \sim U(0, 1)$ \\
$v_{\perp 1} \leftarrow \theta \sqrt{ X_1 } ~\sqrt{\dfrac{X_2}{N^2+X_2}} \cos(2\pi U)$ \\
$v_{\perp 2} \leftarrow \theta \sqrt{ X_1 } ~\sqrt{\dfrac{X_2}{N^2+X_2}} \sin(2\pi U)$ \\
$v_{\parallel}~ \leftarrow \theta \sqrt{ X_1 } ~\dfrac{N}{\sqrt{N^2+X_2} }$ \\
{\bf return} $v_{\perp 1}, v_{\perp 2}, v_{\parallel}$\\
\hline\\
\end{tabular}
\end{minipage}
\begin{minipage}{0.45\textwidth}
\begin{tabular}{l}
\hline
{\bf Algorithm 5.4: KLC distribution}\\
\hline
generate $N \sim \mathcal{N}(0,1)$ \\
generate $Y \sim {\rm Ga}(\kappa - 1/2, 2)$ \\
generate $X_1 \sim {\rm Ga}(3/2,2)$ \\
generate $X_2 \sim {\rm Ga}(j+1,2)$ \\
generate $U \sim U(0, 1)$ \\
$v_{\perp 1} \leftarrow \theta \sqrt{ \dfrac{ \kappa X_1  }{ Y }} \sqrt{\dfrac{X_2}{N^2+X_2}} \cos(2\pi U)$ \\
$v_{\perp 2} \leftarrow \theta \sqrt{ \dfrac{ \kappa X_1  }{ Y }} \sqrt{\dfrac{X_2}{N^2+X_2}} \sin(2\pi U)$ \\
$v_{\parallel}~ \leftarrow \theta \sqrt{ \dfrac{ \kappa X_1  }{ Y }} \dfrac{N}{\sqrt{N^2+X_2} }$ \\
{\bf return} $v_{\perp 1}, v_{\perp 2}, v_{\parallel}$\\
\hline
\end{tabular}
\end{minipage}
\end{table}

\subsection{Loss-cone and kappa loss-cone distributions}
\label{sec:application}

Here, we show practical applications of the loss-cone transform method. 
We first discuss the following PA-type loss-cone distribution.
This one is based on the isotropic Maxwellian.
\begin{align}
f(\vec{v}) &=
\frac{N_0}{\pi^2\theta^3}
\frac{2 \Gamma(j+3/2)}{\Gamma(j+1)}
\Big( \frac{v_{\perp}}{v} \Big)^{2j} \exp\Big( - \frac{v^2}{\theta^2} \Big)
\label{eq:LC-PA}
\\
P_\parallel &=
\frac{3}{2}
~
\frac{1}{2j+3}
N_0 m\theta^2, ~~~
P_\perp =
\frac{3}{2}
~
\frac{j+1}{2j+3}
N_0 m\theta^2
\end{align}
Note that the normalization constant in Eq.~\eqref{eq:LC-PA} is
rescaled by Eq.~\eqref{eq:W}.
It looks complicated, but it usually does not appear in algorithms.
In Section II in ZN22,
the authors discussed that $|v|$ in the Maxwellian follows
the gamma distribution with shape $k=3/2$.
\begin{align}
f(v) &\propto 4\pi v^2 \exp \Big( -\frac{ v^2 }{\theta^2} \Big)
\sim {\rm Ga}(3/2,1)
\end{align}
Then $v$ is given by a gamma-distributed variate.
\begin{align}
v = \theta \sqrt{ X_{{\rm Ga}(3/2,1)} }
\end{align}
After generating $v$ using a gamma-distribution generator,
we employ the loss-cone transform method in Section \ref{sec:LCT}
to obtain the PA-type loss-cone distribution.
The entire procedure is summarized in Algorithm 5.3 in Table \ref{table:LC-PA}.

Finally, we similarly construct a procedure for
the PA-type kappa loss-cone (KLC) distribution. 
This form is consistent with one in an earlier literature.\citep{xiao98}
\begin{align}
f(\vec{v})
&= \frac{N_0}{\pi^2\theta^3\kappa^{3/2}}
\frac{2\Gamma(j+3/2)\Gamma(\kappa+1)}{\Gamma(j+1)\Gamma(\kappa-1/2)}
\nonumber \\
&~~~
\times
\Big( \frac{v_{\perp}}{v} \Big)^{2j}
\Big( 1 + \frac{ v^2 }{\kappa \theta^2} \Big)^{-(\kappa+1)}
\label{eq:KLC-PA}
\\
P_{\parallel} &=
\frac{3\kappa}{2\kappa-3}
~
\frac{1}{2j+3}
N_0 m \theta^2
,~
P_{\perp} = 
\frac{3\kappa}{2\kappa-3}
~
\frac{j+1}{2j+3}
N_0 m \theta^2
\end{align}
In Section III in ZN22,
the authors have shown that
the $|v|$ distribution of the kappa distribution
\begin{align}
\label{eq:kappa_v}
f(v)
& ~\propto~ 4\pi v^2 \Big( 1 + \frac{ {v}^2 }{\kappa \theta^2} \Big)^{-(\kappa+1)} 
\end{align}
is obtained from two gamma-distributed variates.
\begin{align}
v =
\left(
\kappa \theta^2
\frac{
X_{{\rm Ga}(3/2,2)}
}{
X_{{\rm Ga}(\kappa-1/2,2)}
}
\right)^{1/2}
\label{eq:v_recovery}
\end{align}
Combining this with the loss-cone transform method,
we generate the PA-type KLC distribution.
The entire procedure is summarized in Algorithm 5.4 in Table \ref{table:LC-PA}.

\begin{figure}[htbp]
\centering
\includegraphics[width={\columnwidth}]{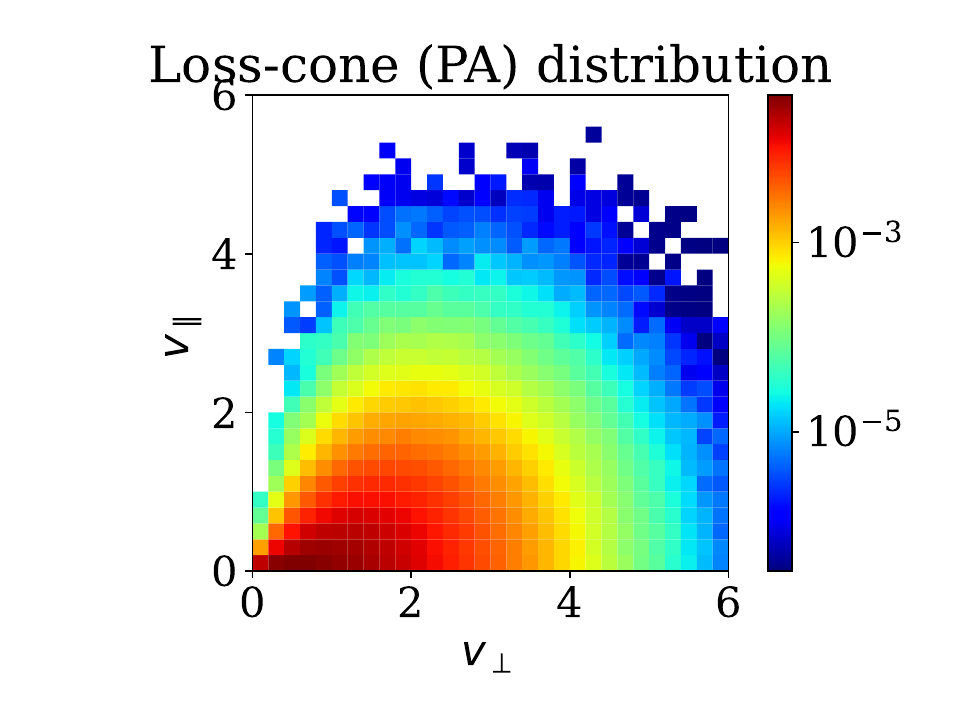}
\caption{
Monte Carlo sampling of the pitch angle (PA)-type loss-cone distribution ($\theta=2.0$ and $j=2.0$) with $10^6$ particles.
Phase-space density is shown in $v_{\perp}$--$v_{\parallel}$.
\label{fig:LC-PA}}
\includegraphics[width={\columnwidth}]{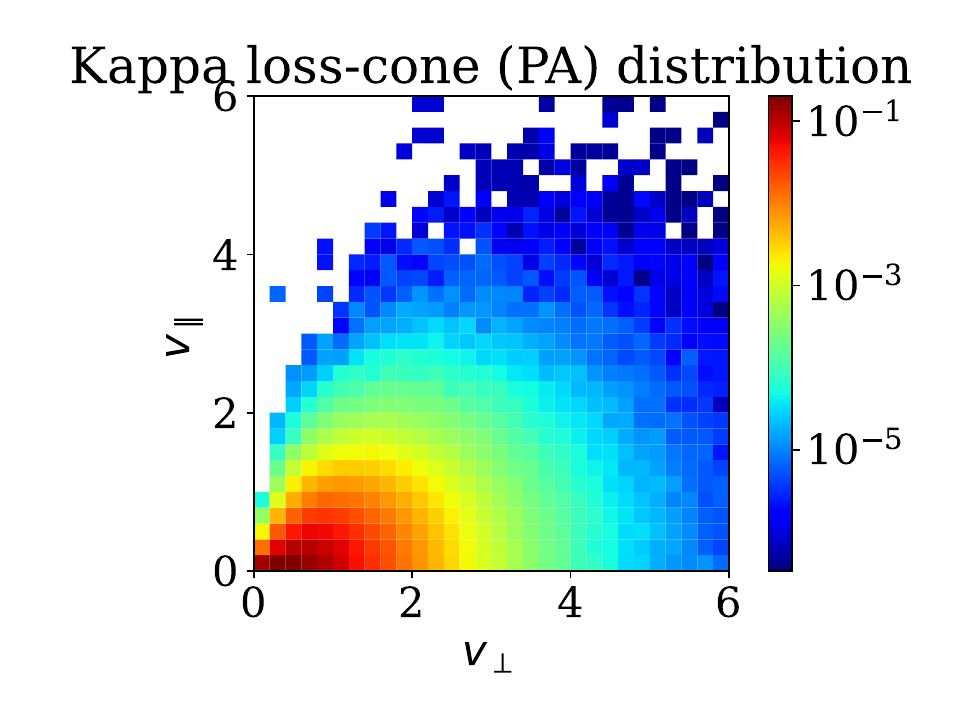}
\caption{
Monte Carlo sampling of the pitch angle (PA)-type KLC distribution ($\theta=1.0$, $j=2.0$, and $\kappa=3.5$) with $10^6$ particles.
Phase-space density is shown in $v_{\perp}$--$v_{\parallel}$.
\label{fig:KLC-PA}}
\end{figure}

Using $10^6$ particles, we have numerically generated
the PA-type loss-cone distribution with $\theta=2$ and $j=2$ and
the PA-type KLC distribution with $\theta=1.0$, $j=2$, and $\kappa=3.5$,
respectively.
Figs. \ref{fig:LC-PA} and \ref{fig:KLC-PA} show
their phase-space densities in $v_{\perp}$--$v_{\parallel}$
in the same format as other plots. 
The density cavities near the $v_{\parallel}$ axis look very different from
those in the other distributions (Figs.~\ref{fig:AK}, \ref{fig:DGH}, and \ref{fig:KLC}).
It is cone-shaped in the PA-type distributions,
while it looks like a vertical hole in the other distributions.
It is clear that
the PA-type distribution better approximates the loss-cone.

\section{Numerical tests}
\label{sec:test}

\begin{figure}[htbp]
\centering
\includegraphics[width={\columnwidth}]{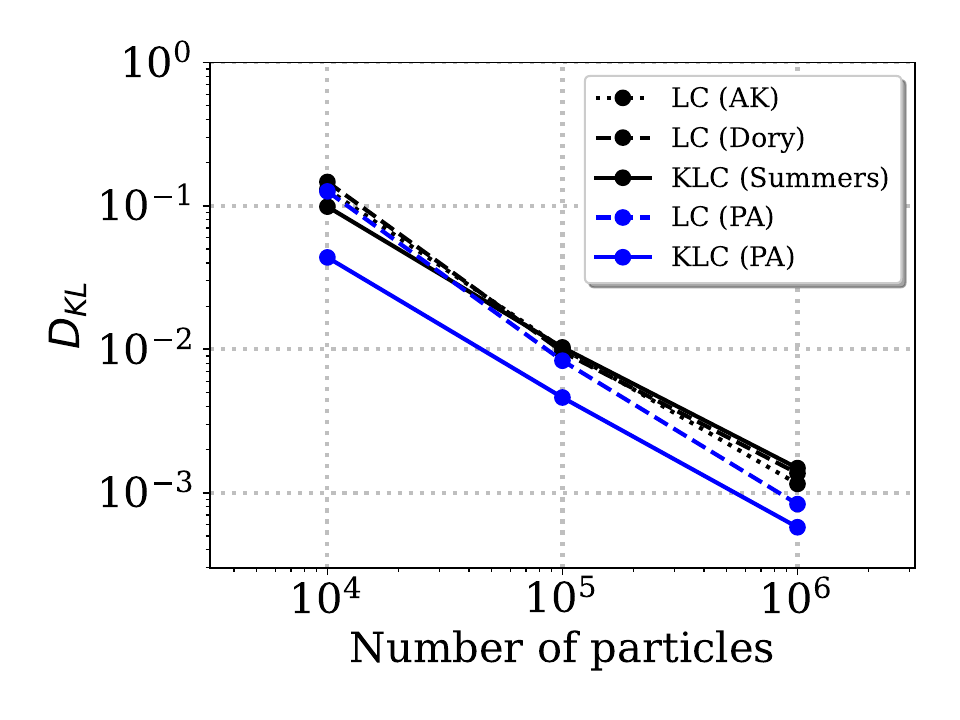}
\caption{
The Kullback--Leibler divergence of the numerical distributions from the analytical solutions (Eq.~\eqref{eq:KL}).
\label{fig:KL}}
\end{figure}

To verify the numerical results for multidimensional velocity distribution functions,
we evaluate the Kullback--Leibler (KL) divergence
between the analytic solutions and our Monte Carlo results.
The KL divergence between two probability distributions, $P$ and $Q$,
are defined by
\begin{align}
D_{\rm{KL}}(P\|Q) = \sum_i P(i) \log \frac{P(i)}{Q(i)}
\end{align}
This quantifies the deviation of $Q$
from the other distribution $P$, and
approaches zero when the two distributions are similar. 
For the target probability distributions $Q(i)$,
we use our Monte Carlo data in the 2-D mesh in the $v_{\perp}$--$v_{\parallel}$ space with $\Delta v =1/5$,
which we have used for our 2-D plots (Figs.~\ref{fig:AK}, \ref{fig:DGH}, \ref{fig:KLC}, \ref{fig:LC-PA}, and \ref{fig:KLC-PA}).
For the baseline probability distributions $P(i)$,
we have numerically integrated the analytic solution by using the same mesh. 
In practice,
we use a very small number $\epsilon=10^{-10}$ to avoid $\log 0$,
\begin{align}
D_{\rm{KL}}(P\|Q) \equiv \sum_i (P(i)+\epsilon) \log \frac{P(i)+\epsilon}{Q(i)+\epsilon}
\label{eq:KL}
\end{align}

Fig.~\ref{fig:KL} shows $D_{\rm{KL}}$ for all the multidimensional velocity distributions,
as a function of the particle numbers.
The absolute value of the KL divergence is unimportant,
because it is even affected by the seed of the random numbers.
Here, it is important to see that the KL divergence approaches zero,
as we increase the particle number. 
This tells us that the Monte Carlo algorithms successfully generate
the velocity distributions.

\section{Discussion and Summary}
\label{sec:discussion}

In this article, we have proposed a series of numerical procedures
that generate loss-cone distributions in particle simulations. 
Since many previous authors may have used the acceptance-rejection method,
we discuss key differences between
the acceptance-rejection method and the proposed methods.

The acceptance-rejection method needs a good sampling distribution.
Sometimes we combine multiple sampling distributions,
for example, the uniform distribution near the maximum and
the power-law distribution for the high-energy tail.
In such a case, the program has multiple logical branches.
Then, for each particle, the program has a rejection loop.
Comparing the target distribution and the sampling distribution,
we accept or reject the particle at some probability.
Since the acceptance rate is less than 100\%,
we need to repeat the same procedure again and again, until accepted.
Technically, the program requires
several random variates and conditional statements (if-else statements)
to switch the branches and to accept/reject the particle inside the loop.
The number of iterations is not fixed.

Most of the proposed methods
(in Sections \ref{sec:AK}--\ref{sec:KLC}, \ref{sec:LCT}, and \ref{sec:application})
generate the loss-cone distribution functions
from uniform, normal, and gamma random variates.
The key features, such as the vertical or cone-shaped density hole and
the power-law tail of the KLC distributions,
are reproduced by the combinations of the variates.
As can be seen in Tables, the procedures are simple.
The algorithms do not have logical branches or loops.
They do not even contain a conditional statement.
Since we do not need to repeat the procedure, the number of operations is fixed.
There features are favorable for parallel computing on GPUs and SIMD processors.
We can take full advantage of these processors,
when we execute the same instruction for multiple data.

The computational cost often depends on the total number of random variates.
The proposed methods require a small number of random variates. 
For example, for the subtracted Maxwellian,
Algorithm 2 requires two variates per particle (Eq.~\eqref{eq:new})
to obtain the $v_{\perp}$ (or $x$) distribution.
In contrast, the acceptance-rejection method typically need two or more variates --- one or more to distribute $v_{\perp}$ (or $x$) and another one to accept or reject the particle.
As the acceptance rate is below 100\%, some more will be necessary.
Therefore, Algorithm 2 appears to be the best choice for the subtracted Maxwellian.
For other distributions, our methods often rely on the gamma random generator.
The costs depends on one's choices of the gamma generators.
The two generators in Section \ref{sec:DGH} (Eqs.~\eqref{eq:gamma_int} and \eqref{eq:gamma_half})
use multiple random variates, in particular when the shape parameter $k$ is large.
In case they are slow, one may try the \citet{mt00} method instead,
which is relatively insensitive to $k$. 

In the Dory-type and PA-type loss-cone distributions and in the KLC distributions,
the loss-cone index $j$ controls the opening angle of the loss-cone.
In the KLC distributions,
the kappa index $\kappa$ controls the spectral index of the power-law tail.
We emphasize that these indices are not limited to integers.
They are arbitrary in the range $j \ge 0$ and $\kappa > 3/2$.
For example, one can set $j=1/2$ or even less to obtain a narrow loss-cone.
In practice, we can just use the gamma generators for half- or non-integer shape parameters, as outlined in Section \ref{sec:DGH} (Appendix A in ZN22 also provides a quick summary).


To construct the PA-type loss-cone distributions,
we started from isotropic base distributions.
We can also consider anisotropic variants of the PA-type distributions.
We simply need to replace
\begin{align}
\frac{1}{\theta^3}
\rightarrow \frac{1}{\theta_{\parallel} \theta_{\perp}^2}
,~~
\frac{v^2}{\theta^2}
\rightarrow 
\frac{v_{\parallel}^2}{\theta_{\parallel}^2} +
\frac{v_{\perp}^2}{\theta_{\perp}^2}
\\
\Big( \frac{v_{\perp}}{v} \Big)^{2j}
\rightarrow
\left( \dfrac{ {v_{\perp}^2}/{\theta_{\perp}^2} }
{{v_{\parallel}^2}/{\theta_{\parallel}^2} + {v_{\perp}^2}/{\theta_{\perp}^2}} \right)^{j}
\label{eq:losscone_anisotropic}
\end{align}
in Eqs.~\eqref{eq:LC-PA} or \eqref{eq:KLC-PA}, and then
$\theta \rightarrow \theta_{\perp}$ for $v_{\perp 1}, v_{\perp 2}$ and
$\theta \rightarrow \theta_{\parallel}$ for $v_{\parallel}$
in algorithms in Table \ref{table:LC-PA}.
Meanwhile, the fractional equation in parentheses in Eq.~\eqref{eq:losscone_anisotropic} no longer mean the sine square of the pitch angle.

The loss-cone transform methods are also applicable to
relativistic velocity distributions.
In fact, the authors have presented
numerical procedures to load
a relativistic Maxwell distribution and
a relativistic kappa distribution in ZN22.
In these cases, we generate
the radial profile of the relativistic four-velocity $\vec{u} = \gamma \vec{v} = [1-(v/c)^2]^{-1/2}\vec{v}$.
Thus, one can straightforwardly combine
these algorithms and the loss-corn transform methods,
to generate relativistic loss-cone and relativistic KLC distributions.\citep{xiao98}

In summary, we have presented Monte Carlo algorithms to
generate loss-cone distributions,
the subtracted Maxwellian (AK-type loss-cone distribution),
the Dory-type loss-cone distribution, and
the Summers-type KLC distributions.
We have further presented another family of distributions,
the PA-type loss-cone distributions.
We have proposed the loss-cone transform methods,
one of which are applied to the PA-type loss-cone and KLC distributions.
Numerical recipes for all these distributions and
the transform methods are provided in Tables.
With some help from gamma-distribution generators,
these algorithms can be easily implemented in one's own code.

\begin{acknowledgements}
We thank Rumi Nakamura for comments on the manuscript.
This work was supported by Grant-in-Aid for Scientific Research (C) 21K03627 and (S) 17H06140 from the Japan Society for the Promotion of Science (JSPS).
\end{acknowledgements}

\section*{Author Declarations}
\subsection*{Conflict of Interest}
The authors have no conflicts to disclose.

\subsection*{Author Contributions}
{\bf Seiji Zenitani}: Conceptualization (lead), Formal analysis (equal), Investigation (lead), Methodology (lead), Visualization (lead), Writing – original draft (lead), Writing – review \& editing (equal), 
{\bf Shin'ya Nakano}: Formal analysis (equal), Methodology (supporting), Writing – review \& editing (equal)

\subsection*{Data Availability Statement}
The Jupyter notebook to generate the figures is available at {\tt arXiv:2309.06879} as an Ancillary file.

\end{document}